\documentclass[twocolumn,prb,aps,superscriptaddress,showpamps]
{revtex4-2}
\usepackage{amsmath,amssymb,bm,graphicx}
\usepackage[hidelinks]{hyperref}
\usepackage{float}
\usepackage{orcidlink}
\usepackage[normalem]{ulem}
\newcommand{\be}{\begin{equation}}

\newcommand{\ee}{\end{equation}}
\newcommand{\bea}{\begin{eqnarray}}
\newcommand{\eea}{\end{eqnarray}}

\renewcommand{\vec}[1]{\mathbf{ #1}}

\renewcommand{\tilde}{\widetilde}

\def\nn{\nonumber\\}

\newcommand{\addEH}[1]{{\color{black}{#1}}}

\newcommand{\addLK}[1]{\textcolor{black}{#1}}

\begin{document}

\title{Dynamic moir\'e-like band modulation in Dirac materials via multi-beam optical interference}

\author{Evelyn P. Sinaga} 
\email{evelyn.sinaga@matanauniversity.ac.id}
\affiliation{Department of Physics, Matana University, South Tangerang 15810, Indonesia}

\author{Rizky Setiawan}
\affiliation{Department of Physics, Airlangga University, Surabaya 60115, Indonesia}

\author{Herri Trilaksana}
\affiliation{Department of Physics, Airlangga University, Surabaya 60115, Indonesia}

\author{Lukas P. A. Krisna}
\affiliation{Research Center for Quantum Physics, National Research and Innovation Agency (BRIN), South Tangerang 15314, Indonesia}

\author{Eddwi H. Hasdeo}
\email{eddw001@brin.go.id}
\affiliation{Research Center for Quantum Physics, National Research and Innovation Agency (BRIN), South Tangerang 15314, Indonesia}

\begin{abstract}
We propose a purely optical Floquet framework to dynamically generate moir\'{e}-like superlattices and quasicrystalline potentials in a gapped Dirac material, entirely bypassing the need for physical twisting. The interference of three coherent circularly polarized beams produces a triangular, valley-dependent Floquet mass landscape that folds the Dirac spectrum into a mini-Brillouin zone, driving pronounced valley-selective miniband reconstruction. We further show that the bands near the Fermi level undergo a light-driven topological phase transition. Remarkably, interfering five or seven beams yields non-crystallographic spatial patterns, giving rise to optical quasicrystals with similar valley contrasting features. Although this purely optical approach provides a versatile route to band engineering, reaching deep subwavelength modulation requires overcoming the free-space diffraction limit. To overcome this limitation, we consider highly confined, linearly polarized surface plasmon polariton fields with controlled relative phases. Their phase-controlled interference generates a spatially alternating chiral mass with zero spatial average. Consequently, the two valleys exhibit degenerate energy spectra while hosting opposite valley-resolved miniband Chern numbers. Our findings establish a reconfigurable, all-optical platform for dynamically engineering moir\'{e}-like bands, quasicrystalline electronic states, and valley-dependent topology without mechanical twisting.

\end{abstract}

\maketitle

\section{Introduction}

The discovery of moir\'e superlattices \addLK{in twisted van der Waals systems} has revolutionized the study of two-dimensional (2D) materials by providing a highly tunable platform for \addLK{exploring strongly correlated physics, non-trivial band topology, and emergent quantum phenomena.}
\cite{Andrei2021Marvels,Cao2018Unconventional,Cao2018Correlated,Wang2020Correlated,Xu2020Correlated,Li2021Quantum,Cai2023Signatures,Guo2025Superconductivity}. 
However, relying on mechanical twisting to achieve the exact ``magic-angle'' configuration presents a 
thermodynamic bottleneck. 
\addLK{During the fabrication processes, local twist-angle variations and inhomogeneous heterostrains spontaneously develop across the sample
\cite{Yoo2019Reconstruction,Jiang2019Charge,Uri2020Mapping}. 
These uncontrolled structural disorders severely distort the narrow flat-band manifold, break crystalline symmetries unpredictably, and smear out the fragile correlated states \cite{Uri2020Mapping,Wilson2020Disorder,Nakatsuji2022Moire}}. 
\addLK{One alternative} to circumvent the difficulties of physical twisting is to place the 2D materials on patterned dielectric superlattices to induce spatial modulations without twisting \cite{Forsythe2018Patterned,Zhan2025Designing}.
\addLK{While promising, these structures remain fixed once fabricated.
Consequently, realizing a pristine, spatially coherent superlattice potential with continuous and in-situ reconfigurability remains a significant experimental challenge.}

\addLK{
An alternative to static superlattices is through Floquet engineering, 
where a time-periodic electromagnetic driving dynamically dresses electronic states and renormalizes the band structure without physical modification of the crystal lattice \cite{Oka2009Photovoltaic,Oka2019Floquet,Rudner2020Band,Bao2022Light}.
In 2D Dirac systems, off-resonant circularly polarized (CP) light explicitly breaks time-reversal symmetry. 
Virtual photon absorption and emission processes generate a non-trivial, optical-chirality-dependent commutator term in the high-frequency Floquet Hamiltonian which opens a topological mass gap at the Dirac points \cite{Kitagawa2011Transport,Ezawa2013Photoinduced,Bukov2015Universal}.
When coupled to systems with broken inversion symmetry or an intrinsic mass gap, the optically synthesized mass interferes valley-dependently with the native mass term, shifting the $K$ and $K^{\prime}$ band edges asymmetrically and enabling valley-selective topological phase transitions \cite{Ezawa2013Photoinduced,Kibis2017All,Mitra2024Light}.
These dynamic band-tailoring principles have been experimentally verified through the observation of photon-dressed Floquet–Bloch states \cite{Wang2013Observation,Mahmood2016Selective,Merboldt2025Observation,Choi2025Observation,Wang2026Observation}, 
and the realization of light-induced anomalous Hall transport in monolayer graphene \cite{Mciver2020Light}.
Despite these successes, standard Floquet schemes typically apply spatially uniform optical fields, modulating the band structure homogeneously. 
Translating this optical control into a spatially-varying modulations \cite{Calvo2025Engineering} offers an unexplored route toward reconfigurable, moir\'e-like band engineering entirely free of mechanical disorder.
}

In this paper, we propose a design of moir\'e-like optical potential entirely out of light. The interference of three CP plane waves intersecting at $120^\circ$ naturally creates a triangular effective mass landscape. Furthermore, generalizing this interference to five or seven beams breaks translational symmetry. This provides a unique, purely optical pathway to generate Penrose-like quasicrystals \cite{Ahn2018Dirac,Yao2018Quasicrystalline} on a single, untwisted monolayer, enabling dynamic control over the band topology.

While this purely optical approach provides a versatile framework for band engineering, realizing such superlattices faces a fundamental experimental roadblock. 
As the superlattice lengthscale is completely determined by light wavelength, free photon source would be irrelevant for a typical electronic lengthscale which is in the order of 10 nm. 
The standard nanophotonic workaround is to utilize surface plasmon polaritons (SPPs) or slow light, which easily achieve 10 nm sub-diffraction confinement \cite{Gramotnev2010Plasmonics,Koppens2011Graphene}. 
However, this introduces a critical physical constraint: plasmons are strictly longitudinal, linearly polarized (LP) modes \cite{Maier2007Plasmonics}. Because they lack intrinsic chirality, interfering with them preserves time-reversal symmetry and fundamentally fails to open the required topological Floquet gap.

\addLK{Alternatively,} we also propose a strategy to synthesize a nanoscale superlattice of CP light by introducing a controlled phase delay between interfering LP modes from plasmons. 
We mathematically demonstrate that this phase delay synthesizes robust local in-plane optical chirality, dynamically generating a deeply modulated topological mass landscape. 
This framework circumvents both the optical diffraction limit and the near-field polarization constraint. 
By providing a highly feasible optical architecture to dynamically dial into scalable topological phases, our approach bypasses the thermodynamic instability of physically twisted layers.

The remainder of this paper is organized as follows. In Sec. II, we introduce the Floquet effective Hamiltonian driven by CP light. In Sec. III, we expand the formalism to multi-beam interference, demonstrating the emergence of quasicrystalline spectra. Sec. IV details the phase-delayed plasmonic interference mechanism and proves the synthesis of local optical chirality. Finally, we summarize our findings and discuss experimental feasibility in Sec. V.

\section{Floquet Theory in Gapped Dirac Materials}
\label{sec:2-cp}

We now construct the theoretical framework for generating a moir\'e-like optical potential purely through electromagnetic means. This approach conceptually parallels the band flattening observed in patterned dielectrics \cite{Forsythe2018Patterned,Zhan2025Designing,Li-kun2020gate}, but operates dynamically by utilizing the interference of multiple coherent laser beams to synthesize a spatially periodic optical superlattice.

Before constructing the spatially modulated optical potentials, it is instructive to establish the Floquet mechanism for a uniform, normal incident electromagnetic field. We consider the low-energy Hamiltonian for electrons around the Dirac points ($K$ and $K^{\prime}$ valleys) in a two-dimensional gapped Dirac material:
\begin{equation}
    H_0^\tau(\vec p) = v(\tau p_x \sigma_x + p_y \sigma_y) + \Delta \sigma_z,
\end{equation}
where $v$ is the Fermi velocity, $p_{x,y}$ are the electron momentum components, $\sigma_{x,y,z}$ are the Pauli matrices representing the sublattice pseudospin, $\Delta$ is the static mass gap (half the total bandgap), and $\tau = \pm 1$ is the valley index for the $K$ and $K^{\prime}$ valleys, respectively.

\subsection{Dynamical Bandgap Modification via Circularly Polarized Light}

We subject the material to a single beam of normal incident, circularly polarized (CP) light, characterized by a time-dependent vector potential:
\begin{equation}
    \mathbf{A}(t) = A_0 (\cos(\omega t)\hat{x} + \eta \sin(\omega t)\hat{y}),
\end{equation}
where $\omega$ is the driving frequency, $A_0$ is the field amplitude, and $\eta = \pm 1$ dictates the optical chirality ($\eta = +1$ for left-circular and $\eta = -1$ for right-circular polarization). 

The light-matter interaction is incorporated via the Peierls substitution, $\mathbf{p} \rightarrow \mathbf{p} - e\mathbf{A}(t)$, which yields a time-dependent interaction Hamiltonian
\begin{equation}
    H_{\text{int}}^\tau(t) = -ev (\tau A_x(t) \sigma_x + A_y(t) \sigma_y).
\end{equation}
Decomposing this into its Fourier harmonics, we isolate the first-harmonic $m=+1$ and $m=-1$ components:
\begin{align}
    H_{\pm1}^\tau &= -\frac{ev A_0}{2} (\tau \sigma_x \pm i \eta \sigma_y).
\end{align}

In the high-frequency regime ($\hbar\omega \gg \Delta, E_F$), the stroboscopic dynamics of the driven system can be described by a static effective Hamiltonian derived from the first-order Magnus expansion (see Appendix~\ref{sec:gfactor})~\cite{Magnus1954Exponential,Kitagawa2011Transport}:
\begin{equation}
    H_{\text{eff}}^\tau \approx H_0^\tau + \frac{[H_{-1}^\tau, H_1^\tau]}{\hbar\omega} + \mathcal{O}\left(\frac{1}{\omega^2}\right),
\end{equation}
where $\mathcal{O}(\omega^{-2})$ contains higher order nested commutators.
Evaluating the first-order commutator explicitly yields $[H_{-1}^\tau, H_1^\tau] = -e^2 v^2 A_0^2 \tau \eta \sigma_z$. Substituting this back into the effective Hamiltonian demonstrates that the CP light acts purely as a mass term, directly modifying the energy gap:
\begin{equation}
    H_{\text{eff}}^\tau = v(\tau p_x \sigma_x + p_y \sigma_y) + \left( \Delta - \frac{(ev A_0)^2}{\hbar\omega}\tau\eta \right) \sigma_z.
\end{equation}
The effective, photon-dressed mass is therefore $\Delta_{\text{eff}}^\tau = \Delta - \frac{(ev A_0)^2}{\hbar\omega}\tau\eta$. Because this dynamical correction is proportional to the valley index $\tau$, the degeneracy between the $K$ and $K^{\prime}$ valleys is explicitly broken \addLK{as long as the intrinsic gap $\Delta$ remains finite}. 
The comparison of light intensity with the gap splitting at two valleys is a gigantic valley $g$ factor whose value is reaching 45~[see~\ref{sec:gfactor}].

Furthermore, it is instructive to contrast the high-frequency Magnus-Floquet framework utilized in this study with the near-resonant (sub-gap) driving regime demonstrated by Sie et al.~\cite{Sie2015Valley}. In the near-resonant regime, the incident photon energy is deliberately tuned just below the unperturbed bandgap. Because this sub-gap optical Stark effect is strictly governed by quantum level repulsion across a two-level virtual crossing, it provides only unidirectional gap control. Specifically, the bandgap exclusively enlarges for the particular valley that matches the circular polarization index of the incident light, while the energy spectrum of the opposite valley remains completely intact. (We refer interested readers to Appendix~\ref{sec:resonant_floquet} for a detailed discussion of this sub-gap regime and its foundational connection to the valley-selective optical Stark effect.)

In sharp contrast, under the high-frequency driving conditions modeled here, the dynamically generated Floquet mass acts as a bidirectional synthetic field. Rather than being limited to unidirectional enlargement, the primary bandgap can be selectively driven to become either smaller or larger. Crucially, this bidirectional tuning is what enables the gap closures necessary to drive Floquet topological phase transitions

\begin{figure}
    \centering
    \includegraphics[width=4cm, trim=0.4cm 0 0.4cm 0, clip]{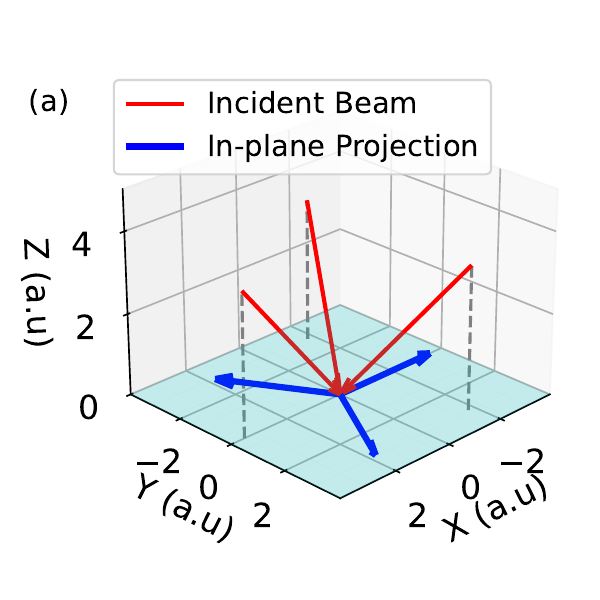}
    \includegraphics[width=4cm]{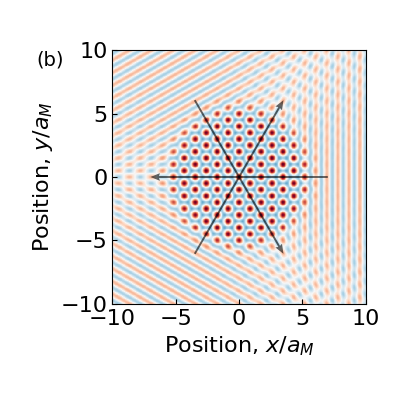}
    \caption{(a) Illustrates the interference geometry: red arrows denote the incident plane waves, and blue arrows denote their in-plane wavevector projections, mutually separated by $120^\circ$. (b) Resulting \addEH{electric field intensity} exhibiting optical triangular superlattice from the interference pattern in the sample plane. $a_M = 2\pi\lambda/3$ is the lattice period, related to the incident light wavelength $\lambda$.}
    \label{fig:3moire}
\end{figure}

\subsection{Optical Superlattice Configuration}

To create a two-dimensional triangular moir\'e-like potential, we employ three coherent, circularly polarized plane waves intersecting at the sample plane. The in-plane wavevectors $\mathbf{q}_1, \mathbf{q}_2,$ and $\mathbf{q}_3$ are arranged symmetrically with $120^\circ$ angular separation, such that $|\mathbf{q}_1| = |\mathbf{q}_2| = |\mathbf{q}_3| = q$.

The total vector potential of this interfering optical field is the superposition of the three beams:
\begin{equation}
    \mathbf{A}(\mathbf{r}, t) = \sum_{j=1}^3 \text{Re} \left[ A_0 (\hat{x} + i\eta\hat{y}) e^{i(\mathbf{q}_j \cdot \mathbf{r} - \omega t)} \right],
\end{equation}
where $\eta = \pm 1$ represents the helicity of the light. Factoring out the time dependence and the polarization vector, we can rewrite this superposition as
\begin{equation}
    \mathbf{A}(\mathbf{r}, t) = \frac{A_0}{2} (\hat{x} + i\eta\hat{y}) f(\mathbf{r}) e^{-i\omega t} + \text{c.c.}
\end{equation}
where the complex spatial modulation function $f(\mathbf{r})$ is defined as $f(\mathbf{r}) = \sum_{j=1}^3 e^{i\mathbf{q}_j \cdot \mathbf{r}}$. The square value of $|f(\vec r)|^2= 3 + 2 \sum_{i=1}^3 \cos(\vec b_i\cdot \vec r),$ where
\begin{align}
    \mathbf{q}_j &= q_0 \left( \cos\theta_j, \sin\theta_j \right), \quad \theta_j = (j-1)\frac{2\pi}{3}, \nn
    \mathbf{b}_j &= \mathbf{q}_j - \mathbf{q}_{j+1}, \quad (\text{with } \mathbf{q}_4 \equiv \mathbf{q}_1).
\end{align}
This results in the spatially modulated effective mass which is given by
\begin{equation}
    \Delta_{\text{eff}}^\tau(\mathbf{r}) = \Delta - \frac{(evA_0)^2}{\hbar\omega} \eta \tau \left[ 3 + 2 \sum_{l=1}^3 \cos(\mathbf{b}_l \cdot \mathbf{r}) \right]. \label{eq:Deltar}
\end{equation}
This spatial modulation acts as a strong optical periodic scattering potential that folds the original Dirac cone into a miniature moir\'{e}-like superlattice Brillouin zone. 

Figure~\ref{fig:3moire}(a) illustrates the interference geometry: three plane waves (red arrows) are incident on the sample, with their in-plane wavevector components (blue arrows) arranged symmetrically at $120^\circ$ to one another. This threefold symmetry is the minimal configuration needed to generate a two-dimensional optical triangular superlattice. Figure~\ref{fig:3moire}(b) shows the \addEH{electric field intensity} of resulting periodic interference pattern in the sample plane, plotted in units of the lattice constant $a_M = 2\pi\lambda/3$. The orange fringes trace the pairwise diffraction pattern between the three beams, while the red dots mark the intensity maxima, which define the emergent triangular superlattice whose period is set by the incident wavelength $\lambda$.

\subsection{Plane-Wave Expansion of the Moir\'e Bandstructure}

Because the spatial modulation $\Delta_{\text{eff}}^\tau(\mathbf{r})$ [Eq.~\eqref{eq:Deltar}] is periodic with respect to the triangular superlattice, we can employ Bloch's theorem and project the Hamiltonian onto a plane-wave basis. We define a uniform mass shift $m_0^\tau$ and a 
modulation amplitude $V_M^\tau$:
\begin{equation}
    m_0^\tau = \Delta - 3 V_M^\tau, \quad V_M^\tau = -V_0 \eta \tau,
\end{equation}
where $V_0=\displaystyle \frac{(evA_0)^2}{\hbar\omega}.$
The effective mass then takes a highly symmetric form across the set of six nearest-neighbor reciprocal lattice vectors $\mathcal{G}_1 = \{\pm \mathbf{b}_1, \pm \mathbf{b}_2, \pm \mathbf{b}_3\}$:
\begin{equation}
    \Delta_{\text{eff}}^\tau(\mathbf{r}) = m_0^\tau + V_M^\tau \sum_{\mathbf{g} \in \mathcal{G}_1} e^{i\mathbf{g} \cdot \mathbf{r}}.
\end{equation}

By multiplying the Schrödinger equation $H_{\text{eff}}^\tau \Psi_{\mathbf{k}}^\tau = E \Psi_{\mathbf{k}}^\tau$ from the left by $e^{-i(\mathbf{k}+\mathbf{G}')\cdot\mathbf{r}}$ and integrating over real space, the continuous differential equation transforms into the central equation:
\begin{equation}
    \sum_{\mathbf{G}} \mathcal{H}_{\mathbf{G}', \mathbf{G}}^\tau (\mathbf{k}) \begin{pmatrix} c_{A, \mathbf{G}} \\ c_{B, \mathbf{G}} \end{pmatrix} = E_{\mathbf{k}} \begin{pmatrix} c_{A, \mathbf{G}'} \\ c_{B, \mathbf{G}'} \end{pmatrix},\label{eq:central}
\end{equation}
where $\mathbf{G} = n \mathbf{b}_1 + m \mathbf{b}_2$, and $n$ and $m$ are integers bounded by the cutoff $|n|, |m| \le n_{\text{max}}$.
The matrix elements $\mathcal{H}_{\mathbf{G}', \mathbf{G}}^\tau (\mathbf{k})$ couple different momentum states. The diagonal kinetic terms ($\mathbf{G}' = \mathbf{G}$) represent the unperturbed kinetic energy evaluated at the shifted momentum $\mathbf{k} + \mathbf{G}$, plus the uniform mass shift:
\begin{equation}
    \mathcal{H}_{\mathbf{G}, \mathbf{G}}^\tau(\mathbf{k}) = \hbar v  
    \Big[\tau (k_x + G_x)\sigma_x + (k_y + G_y)\sigma_y\Big] +m_0^\tau \sigma_z .
    \end{equation}
The off-diagonal scattering terms ($\mathbf{G}' \neq \mathbf{G}$) only occur if the momentum difference exactly matches one of the six primary reciprocal vectors ($\mathbf{G}' - \mathbf{G} = \mathbf{g} \in \mathcal{G}_1$):
\begin{equation}
    \mathcal{H}_{\mathbf{G}+\mathbf{g}, \mathbf{G}}^\tau = V_M^\tau \sigma_z = \begin{pmatrix} V_M^\tau & 0 \\ 0 & -V_M^\tau \end{pmatrix}.
\end{equation}
Diagonalizing this truncated matrix along the high-symmetry path of the miniature superlattice Brillouin zone ($\boldsymbol{\kappa}_- \rightarrow \boldsymbol{\gamma} \rightarrow \boldsymbol{\mu} \rightarrow \boldsymbol{\kappa}_+$) yields the exact energy dispersion $E_{\mathbf{k}}$. 

\begin{figure*}[t]
\centering
\includegraphics[width=\textwidth]{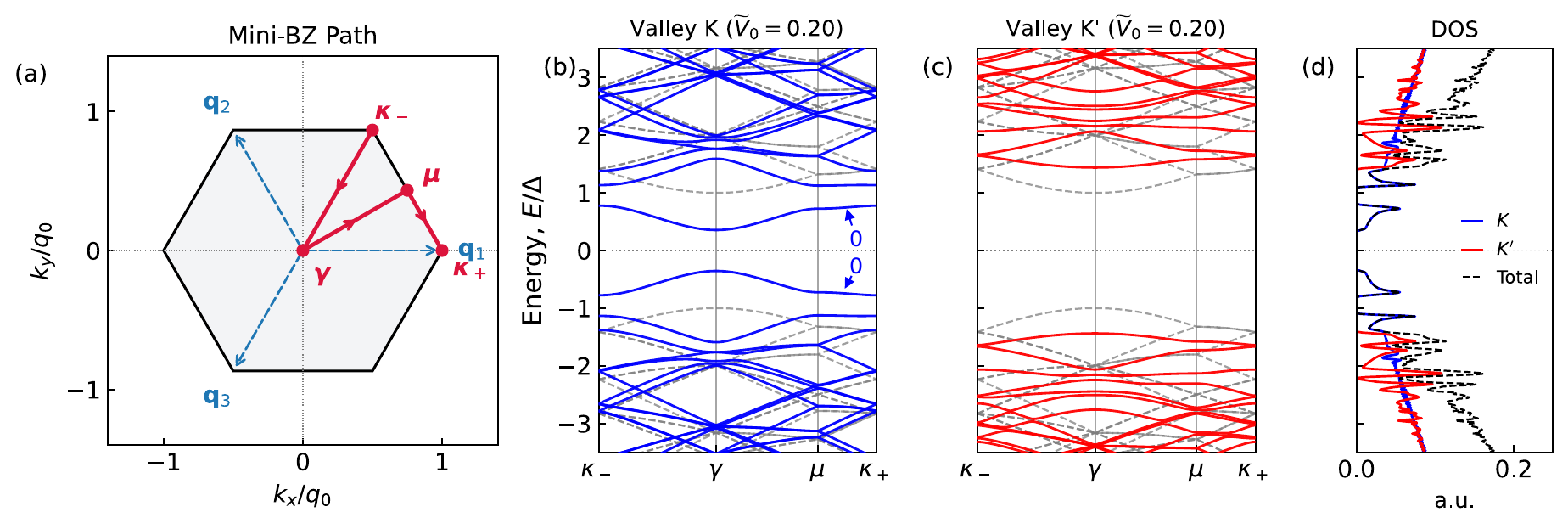}
\caption{(a) The first mini-Brillouin zone (mBZ) of the optical superlattice formed by $N=3$ interfering circularly polarized beams with in-plane wavevectors $\mathbf{q}_1, \mathbf{q}_2, \mathbf{q}_3$ symmetrically separated by $120^\circ$. The red directed path traces the high-symmetry trajectory $\boldsymbol{\kappa}_- \rightarrow \boldsymbol{\gamma} \rightarrow \boldsymbol{\mu} \rightarrow \boldsymbol{\kappa}_+$. (b, c) Valley-resolved moir\'e miniband structures along the  high-symmetry path for (b) valley $K$ ($\tau = +1$) and (c) valley $K'$ ($\tau = -1$). The gray dashed lines indicate the case of $V_0=0$. We used $\hbar v q_0 = \Delta$, and  $\widetilde{V}_0 = V_0/\Delta=0.2$ for the solid lines.
(d) Corresponding valley-resolved and total density of states (DOS) calculated over the full two-dimensional mBZ.}

\label{fig:moire_bands_dos}
\end{figure*}

In Fig.~\ref{fig:moire_bands_dos}(b) and (c) we show the resulting eigenvalues $E_\vec k$ of Eq.~\eqref{eq:central} for valley $K$ and $K'$, respectively. The gray-dashed lines in Fig.~\ref{fig:moire_bands_dos}(b) and (c) depict the folded gapped Dirac band without the moir\'{e}-like potential ($V_0=0$). Here, we find that theoptical CP beams which manifest in $V_0$ lifts the valley degeneracy especially in the low energy range. In the $K$ valley ($\tau=+1$) shown in Fig.~\ref{fig:moire_bands_dos}(b), the destructive interplay between the optical chirality and the valley index results in a reduced effective mass ($m_0^+ = 0.40 \Delta$). Consequently, the Dirac cone folds into the mBZ but remains highly dispersive with a narrowed bandgap, as the electron kinetic energy continues to dominate over the optical scattering potential ($V_0 =  0.20\Delta$).
Conversely, the dynamics within the $K'$ valley ($\tau=-1$) depicted in Fig.~\ref{fig:moire_bands_dos}(c) exhibit a stark contrast. The constructive interference drastically enhances the effective mass ($m_0^- = 1.60\Delta$), opening a large bandgap. Most notably, the lowest-energy isolated bands become remarkably flat across the entire mBZ.

The optical configuration of $N=3$ intersecting beams possesses an intrinsic threefold ($C_3$) spatial symmetry, the resulting real-space mass modulation $\Delta_{\mathrm{eff}}^\tau(\mathbf{r})$ is governed by cosine interference terms, ensuring real-space inversion symmetry $\Delta_{\mathrm{eff}}^\tau(-\mathbf{r}) = \Delta_{\mathrm{eff}}^\tau(\mathbf{r})$. Under the pseudospin conjugation transformation $\sigma_z H_{\mathrm{eff}}^\tau(-\mathbf{k}) \sigma_z = H_{\mathrm{eff}}^\tau(\mathbf{k})$, the scalar energy spectrum acquires an exact inversion symmetry $E^\tau(\mathbf{k}) = E^\tau(-\mathbf{k})$. 
Consequently, the energy spectra at the two inequivalent superlattice corners $\boldsymbol{\kappa}_-$ and $\boldsymbol{\kappa}_+$ are degenerate ($E(\boldsymbol{\kappa}_-) = E(\boldsymbol{\kappa}_+)$), as clearly demonstrated in Figs.~\ref{fig:moire_bands_dos}(b) and \ref{fig:moire_bands_dos}(c).


These spectral features are fundamentally corroborated by the low-energy DOS in Fig.~\ref{fig:moire_bands_dos}(d). The $K$ valley DOS (blue curve) exhibits a continuous distribution of available states at low energies, confirming its dispersive nature. On the contrary, the $K'$ valley DOS (red curve) displays a wide  gap flanked by sharp van Hove singularities at the band edges, which is a signature of flat band spatial localization. 
The resulting total DOS is depicted by the black dashed curve.

\begin{figure}[t]
    \centering
    \includegraphics[trim=0.5cm 1.0cm 0.5cm 1.2cm, width=9cm]{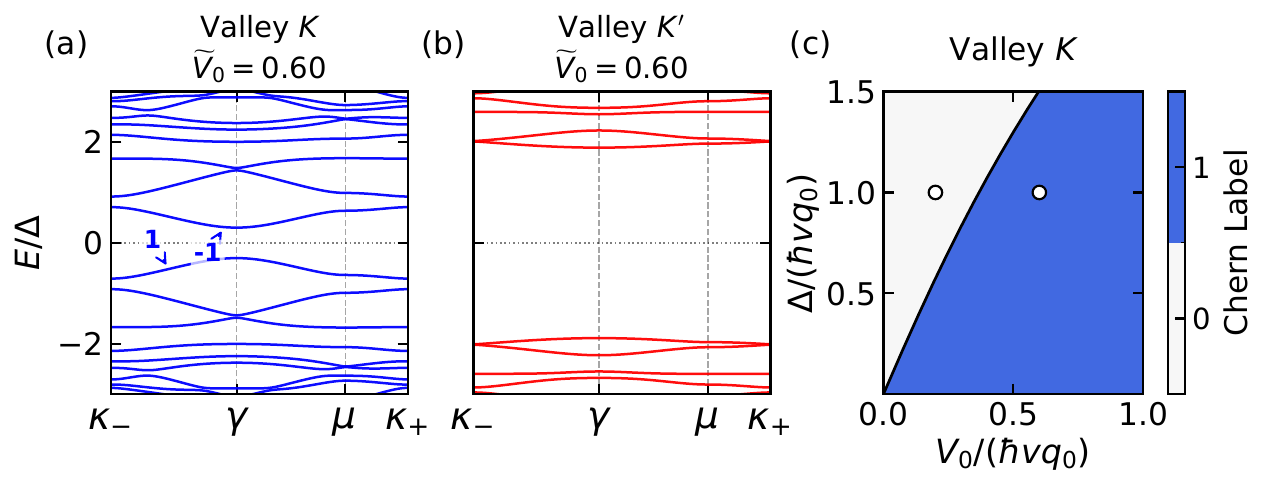}
    \caption{Valley-resolved topological phase transition under circularly polarized light. (a,b) Band structures at $\Delta/(\hbar vq_0)=1.0$ for $V_0/\Delta= 0.60$, representing the $C_v^{K}=1$ for $K$ valley. (c) Corresponding topological phase diagrams for the $K$ valley. Black lines denote the phase boundaries, while the open circles indicate the parameter points used in panels (a) and (b).}
    \label{fig:chern_cpl}
\end{figure}
Computation of valley Chern number reveals an interesting topological phase transition depending on parameters $\Delta$ and $V_0$ in Eq.~\eqref{eq:Deltar}. In Fig.~\ref{fig:moire_bands_dos} (b)  we labeled the Chern numbers of the nearest bands from the Fermi level following the gauge-invariant Fukui-Hatsugai-Suzuki formulation \cite{Fukui2005Chern} [see ~\ref{sec:S5} for details]. As the Chern number is not well-defined in gapless bands, we classify the isolated band using a threshold gap $\approx 0.005 \Delta$ .  The value of Chern numbers of the valence ($v$) and conduction ($c$)  bands near the Fermi level dramatically changes from $0$ to $\pm 1$ upon changing the strength of circular electric field $V_0$ as shown in Fig.~\ref{fig:chern_cpl}(a) .
The corresponding phase diagrams in Figs.~\ref{fig:chern_cpl}(c)  reveal a valley-selective topological phase transition. The $K$-valley $v$ band changes from $C_{v}^{K}=0$ at $V_0/\Delta=0.20$ [Fig.~\ref{fig:moire_bands_dos}(b)] to $C_{v}^{K}=+1$ at $V_0/\Delta=0.60$ [Fig.~\ref{fig:chern_cpl}(a)], accompanied by a redistribution of the Chern numbers of the adjacent minibands from $(C_v^{K},C_c^{K})=(0,0)$ to $(+1,-1)$. This change is consistent with a gap-closing and reopening process associated with a light-driven topological transition \cite{Ezawa2013Photoinduced,jsong15pnas}. A spatially uniform mass picture predicts the topological phase transition at $m_0^K=0$ which corresponds to $V_0=\frac{\Delta}{N}$ with $N=3$. In contrast, the band gap is negligible at valley $K'$ thus excluded in the Chern number assignment.

\section{Optical Quasicrystals from Multi-Beam Interference}

The Floquet band engineering framework is not restricted to periodic optical superlattices. By increasing the number of interfering optical beams, we can break the translational symmetry of the effective mass landscape, providing a purely optical route to realize two-dimensional quasicrystals.

\subsection{Generalization to $N$-Beam Driving and Quasicrystallinity}

We generalize the optical setup to $N$ coherent, circularly polarized plane waves intersecting at the sample plane. The wavevectors are distributed symmetrically in the azimuthal plane:
\begin{equation}
    \mathbf{q}_j = q_0 \left( \cos\left(\frac{2\pi j}{N}\right), \sin\left(\frac{2\pi j}{N}\right) \right), \quad j = 1, 2, \dots, N,\label{eq:multiq}
\end{equation}
where $q_0$ is the characteristic momentum scale of the incident light. The generalized spatial modulation function is $f(\mathbf{r}) = \sum_{j=1}^N e^{i\mathbf{q}_j \cdot \mathbf{r}}$. 

Following the first-order Magnus expansion derived in Sec.~\ref{sec:2-cp}, the local Floquet mass is proportional to the optical intensity $|f(\mathbf{r})|^2$. Expanding this product, the spatially modulated effective mass for an $N$-beam configuration takes the form:
\begin{equation}
    \Delta_{\text{eff}}^\tau(\mathbf{r}) = \Delta -V_0 \eta \tau \left[ N + 2 \sum_{j < k}^N \cos(\mathbf{b}_{jk} \cdot \mathbf{r}) \right].
\end{equation}
where the set of fundamental scattering vectors is defined by all pairwise differences $\mathbf{b}_{jk} = \mathbf{q}_j - \mathbf{q}_k$.

\begin{figure}[t]
    \centering
    \includegraphics[trim=1cm 1cm 3cm 1cm,clip,width=9cm]{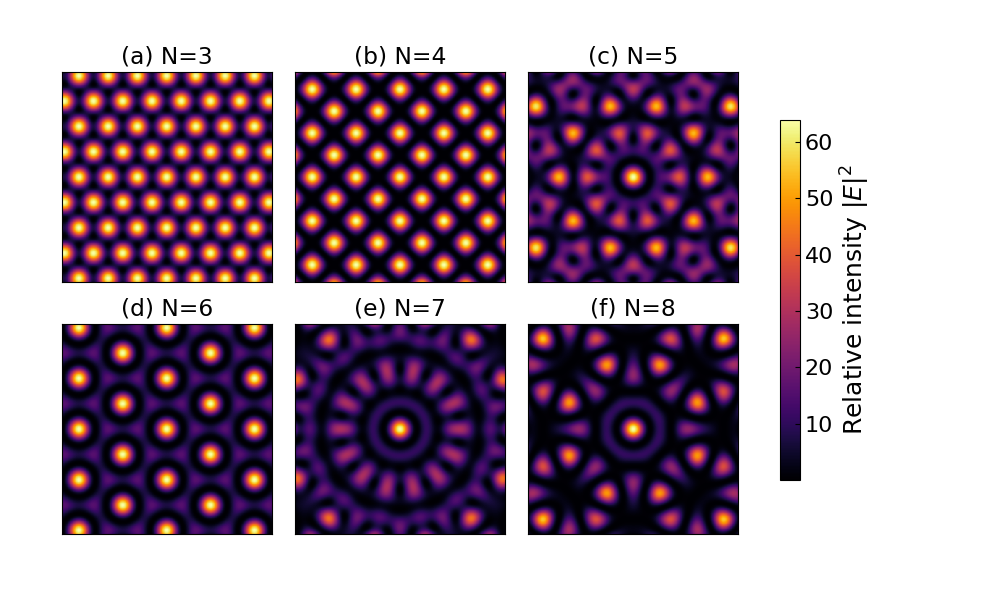}
    \caption{Real-space relative intensity $|E(\mathbf{r})|^2$ of the interference pattern for different numbers of incident beams $N$, according to Eq.~\eqref{eq:multiq}: (a) $N=3$, (b) $N=4$, (c) $N=5$, (d) $N=6$, (e) $N=7$, (f) $N=8$. While $N=3,4,6$ yield periodic moir\'e-like  superlattices, the patterns for $N=5$, $N=7$, and $N=8$ lack translational periodicity and instead show quasicrystalline order.}
    \label{fig:Moire Phase}
\end{figure}

Figure~\ref{fig:Moire Phase} shows the resulting real-space intensity patterns $|E(\mathbf{r})|^2$ for increasing beam number $N$, computed from Eq.~\eqref{eq:multiq}. For $N=3$, $N=4$, and $N=6$ [panels (a), (b), and (d)], the interference pattern remains fully periodic, reproducing triangular, square, and hexagonal moir\'e-like superlattices, respectively. In contrast, for $N=5$, $N=7$, and $N=8$ [panels (c), (e), and (f)], the intensity pattern exhibits five-, seven-, and eightfold rotational symmetry, respectively, none of which is compatible with any two-dimensional periodic lattice. These configurations therefore lack long-range translational order while retaining long-range orientational order, the defining signature of a two-dimensional quasicrystal. For $N=5$ (pentagonal) or $N=7$ (heptagonal), the scattering vectors $\mathbf{b}_{jk}$ cannot be expressed as integer linear combinations of just two primitive reciprocal lattice vectors.  This configuration dynamically imprints a Penrose-like quasicrystalline potential onto the Dirac fermions.

\subsection{Real-Space Evaluation of the Density of States}

Because the $N=5$ and $N=7$ optical potentials are quasicrystalline, Bloch's theorem is rendered invalid. There is no finite Brillouin zone, and the plane-wave expansion utilized for the periodic $N=3$ case leads to an infinitely dense spectrum of momentum states that cannot be smoothly truncated.

\begin{figure}[t]
    \centering
    \includegraphics[width=9cm]{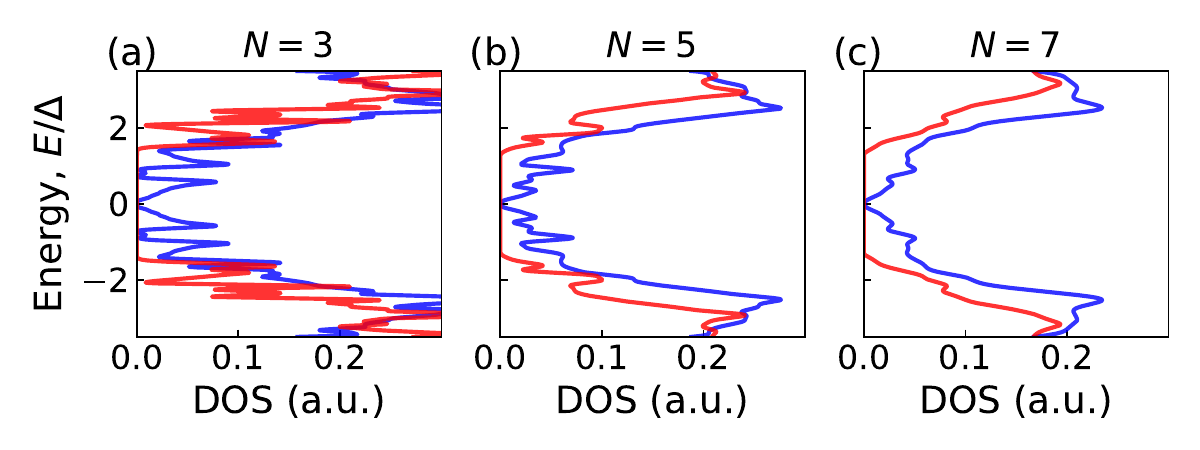}
    \caption{Density of states (DOS) of optical moir\'e  pattern from $N$ laser beams. $N=3$ to be compared with Fig.~\ref{fig:moire_bands_dos}(c). DOS of $N=5$ and $N=7$ represent the electronic structure of quasicrystals.    }
    \label{fig:multibeams_dos}
\end{figure}

To accurately compute the electronic structure of quasicrystalls, we must abandon momentum space and evaluate the Hamiltonian in real space. We map the continuous effective Hamiltonian $H_{\text{eff}}^\tau = v(\tau p_x \sigma_x + p_y \sigma_y) + \Delta_{\text{eff}}^\tau(\mathbf{r}) \sigma_z$, where $p_i = -i \hbar \partial_i$ onto a discrete tight-binding square grid with a lattice spacing $a_0 \ll 2\pi/q_0$.

To handle the macroscopic size of the real-space supercell required to capture the quasicrystalline features without severe finite-size boundary effects, we employ the kernel polynomial method (KPM) \cite{Weisse2006Kernel}. The KPM avoids the computationally prohibitive exact diagonalization of the massive real-space Hamiltonian matrix by expanding the density of states (DOS), $\rho(E)$, in terms of Chebyshev polynomials $T_m(x)$:
\begin{eqnarray}
    \rho(E) &=& \frac{1}{D} \text{Tr}\left[ \delta(E - H_{\text{eff}}^\tau) \right] \nonumber \\
    &\approx& \frac{1}{D \pi \sqrt{1 - \tilde{E}^2}} \sum_{m=0}^{M-1} g_m \mu_m T_m(\tilde{E}),
\end{eqnarray}
where $D$ is the Hilbert space dimension, $\tilde{E}$ is the rescaled energy bound within $[-1, 1]$, and $\mu_m = \text{Tr}[T_m(H_{\text{eff}}^\tau)]$ are the Chebyshev moments computed iteratively via sparse matrix-vector multiplications. The Jackson kernel $g_m$ is introduced to suppress Gibbs oscillations arising from the finite polynomial truncation order $M\approx 800$.

 Figure~\ref{fig:multibeams_dos}(a) is the calculated DOS of the periodic $N=3$ interference pattern. The DOS exhibits highly structured subbands with sharp van Hove singularities and well-defined secondary gaps, demonstrating excellent agreement with the momentum-space band structure  calculations [c.f. Fig.~\ref{fig:moire_bands_dos}(d)]. A pronounced valley splitting is universally observed across all optical configurations; the $K'$ valley (red curves) maintains a significantly larger band gap around $E=0$ compared to the $K$ valley (blue curves). This explicit valley contrast originates from the global Floquet mass shift induced by the circularly polarized driving field, which couples with opposite signs to the two valleys. 

As the number of incident beams increases to generate $N=5$ and $N=7$ quasicrystalline potentials [Figs.~\ref{fig:multibeams_dos}(b) and \ref{fig:multibeams_dos}(c)], the macroscopic translational symmetry of the lattice is explicitly broken. Consequently, the DOS exhibits a distinct $N$-dependent evolution: the sharp, discrete peaks characteristic of the periodic $N=3$ lattice are progressively washed out, resulting in a much smoother, continuous spectral distribution. This spectral smoothing reflects the fragmentation of the electronic bands into the dense, highly degenerate states \cite{Kraus2012Topological}. Despite this structural smearing of the higher-energy subbands, the primary valley-asymmetric band gaps remain robustly preserved, highlighting the stability of the dynamically generated Floquet mass within optical quasicrystals.

\section{Creating Local CP Light from Interferences of LP Light with Phase Delay}

Generating circularly polarized optical fields with a remarkably large wavevector, corresponding to deep sub-wavelength scales on the order of $\lambda \approx 10$ nm, presents a formidable challenge in modern nano-optics. In the free-photon regime, achieving such extreme spatial confinement natively requires utilizing the extreme ultraviolet (EUV) or soft X-ray spectrum. At these high frequencies, natural materials lack the necessary optical birefringence, and standard polarization-control components, such as traditional quarter-wave plates, become highly absorptive and fundamentally ineffective. Consequently, synthesizing a robust, circularly polarized field with large momentum using strictly free-space photons is challenging~\cite{Basov2016Polaritons}.

To overcome this fundamental diffraction limit, one can exploit surface plasmon polaritons (SPPs). Plasmonic waves are collective, coherent oscillations of conduction electrons at a metal-dielectric interface coupled to an electromagnetic field. They naturally exhibit highly confined near-fields with wavevectors significantly larger than those of free-space photons at the same excitation frequency, effectively shrinking the spatial wavelength down to the required $10$ nm regime \cite{Basov2016Polaritons}.

Instead of relying on macroscopic birefringent crystals, which are ineffective at these scales, one can synthesize local circular polarization through the vectorial superposition of multiple intersecting linearly polarized plasmonic beams. By introducing a controlled phase delay between these intersecting beams, one can construct a stationary optical interference pattern, i.e. a polarization lattice, where the local chirality alternates periodically across space.

Consider the interference of $N=3$ coplanar waves. The total electric field $\mathbf{E}(\mathbf{r}, t)$ at a spatial coordinate $\mathbf{r} = (x,y)$ is the vector sum of the individual beams:
\begin{equation}
    \mathbf{E}(\mathbf{r}, t) = \sum_{j=1}^{3} E_0 \hat{\mathbf{n}}_j e^{i(\mathbf{q}_j \cdot \mathbf{r} - \omega t + \phi_j)},
\end{equation}
where $\mathbf{q}_j$ is the in-plane wavevector of the $j$-th beam, $\hat{\mathbf{n}}_j$ is the linear polarization unit vector, and $\phi_j$ represents the initial phase of the beam. Importantly, the phase delay mechanism described here applies universally to any linearly polarized waves. It is not restricted solely to longitudinal plasmonic modes (where $\hat{\mathbf{n}}_j \parallel \mathbf{q}_j$), but is equally valid for transverse free-space modes (where $\hat{\mathbf{n}}_j \cdot \mathbf{q}_j = 0$).

For a symmetric threefold interference, the wavevectors are oriented at angles $\theta_j = (j-1)\frac{2\pi}{3}$. If these beams are perfectly in-phase ($\phi_1 = \phi_2 = \phi_3 = 0$), the electric field components destructively interfere at the intensity maxima, resulting in a strictly linear local polarization state everywhere.

However, if a phase gradient is applied; e.g. $\phi_1 = 0$, $\phi_2 = \frac{2\pi}{3}$, and $\phi_3 = \frac{4\pi}{3}$, the temporal arrival of the wave crests is delayed. At a specific spatial node, the total electric field vector receives sequential contributions from the three beams at different times in the optical cycle, physically forcing the net field vector to rotate in time.
This resulting local chirality is quantified by the degree of circular polarization (DOCP):
\begin{equation}
\text{DOCP} =  \frac{2 \text{Im}(E_x^* E_y)}{|E_x|^2 + |E_y|^2}\label{eq:docp}
\end{equation}
Under the phase-delayed configuration, the DOCP varies spatially across the resulting superlattice. At specific high-intensity nodes, the sequential rotation of the net electric field vector creates a perfectly right-handed circular polarization ($\text{DOCP} = +1$), while adjacent nodes exhibit left-handed circular polarization ($\text{DOCP} = -1$). This spatially alternating chirality acts as a synthetic periodically reversing magnetic field, which is the precise mechanism required to break time-reversal symmetry locally and open topological bandgaps in Floquet-engineered Dirac materials.

\begin{figure}
    \centering
    \includegraphics[trim=1.8cm 0cm 2cm 0cm, clip, width=9cm]{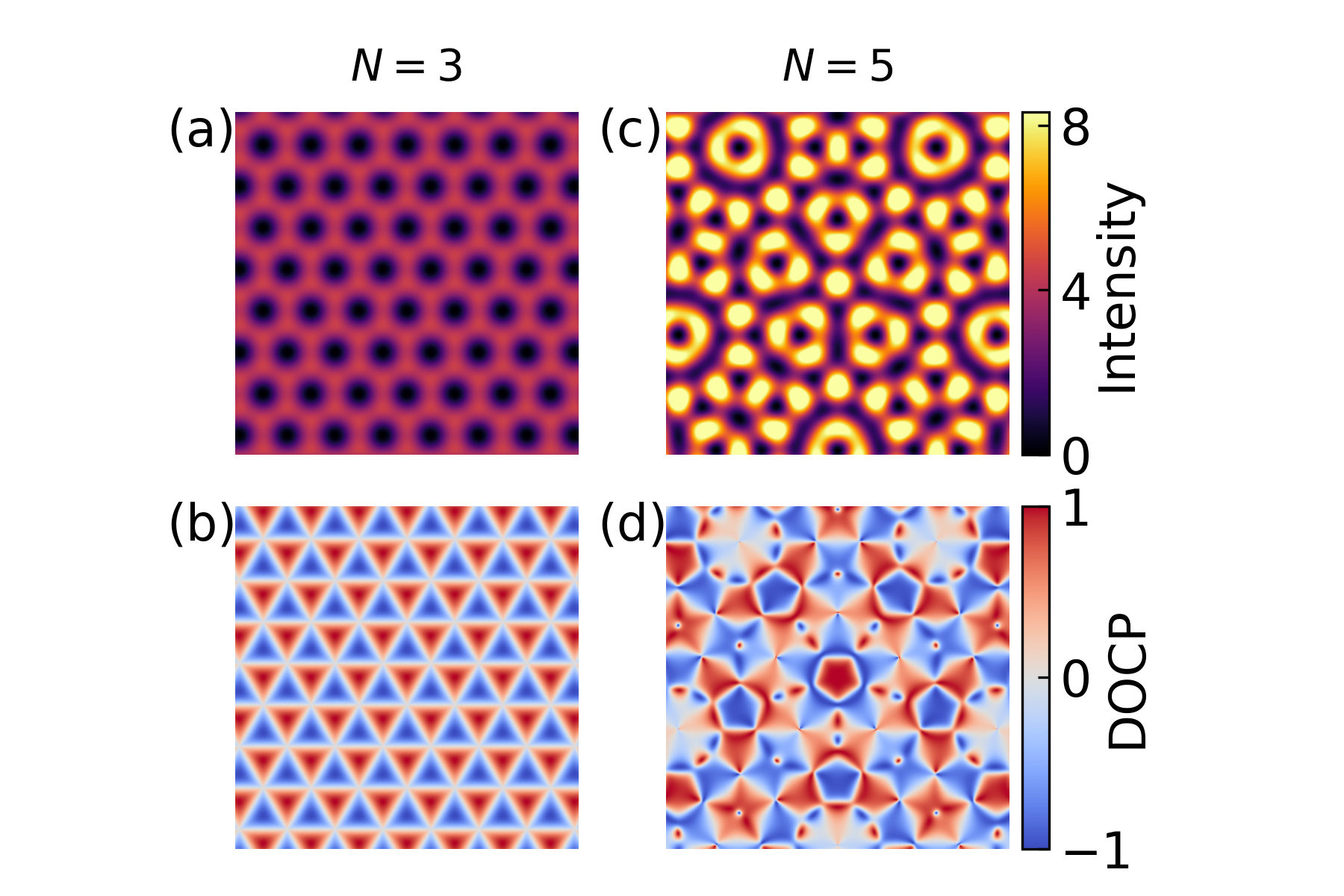}
\caption{Creating CP light from (a,b) $N=3$ and (c,d) $N=5$ beams of LP light with phase delay Eq.~\eqref{eq:Hamlp}. Spatial dependence of (a,b) electric field intensity and (c,d) degree of circular polarization (DOCP) Eq.~\eqref{eq:docp}. \label{fig:phasedelay} }
    \label{fig:lpmoire}
\end{figure}

Figure~\ref{fig:phasedelay} illustrates the spatial profiles of total electric field intensity and the synthesized $\text{DOCP}$ resulting from the phase-delayed interference of $N=3$ and $N=5$ linearly polarized (LP) longitudinal plasmonic beams. For the $N=3$ configuration, the interference of three symmetric beams produces a highly periodic triangular intensity landscape [Fig.~\ref{fig:phasedelay}(a)] with a characteristic spatial period governed by $a_M = 4\pi/(3q_0)=2\lambda/3$. Increasing the beam count to $N=5$ breaks the translational symmetry of the spatial lattice while maintaining long-range 10-fold orientational order [Fig.~\ref{fig:phasedelay}(b)], dynamically imprinting a Penrose-like quasicrystalline potential onto the underlying two-dimensional Dirac material.

Figs.~\ref{fig:phasedelay}(c) and \ref{fig:phasedelay}(d) demonstrate that introducing relative phase delays ($\phi_j = (j-1)2\pi/N$) converts strictly longitudinal LP plasmonic modes into a nanoscale spatial polarization lattice with local in-plane chirality. For $N=3$, the resulting $\text{DOCP}$ landscape forms an alternating triangular array of right-handed ($\text{DOCP} \to +1$, red) and left-handed ($\text{DOCP} \to -1$, blue) circular polarization domains. This spatial variation in optical helicity acts as a synthetic, periodically reversing magnetic field that locally breaks time-reversal symmetry without requiring inherently chiral photon sources or birefringent optics at sub-diffraction scales.

For the quasicrystalline $N=5$ case [Fig.~\ref{fig:phasedelay}(d)], the synthesized optical chirality adopts a complex, self-similar pentagonal pattern where maximal circular polarization ($\text{DOCP} = \pm 1$) is localized at high-symmetry nodes. This proves that phase-delayed plasmonic interference provides a versatile and strictly optical framework to bypass both the optical diffraction limit and near-field polarization constraints. By dynamically generating deeply modulated topological mass landscapes with tuneable spatial symmetries ($N=3, 5$), this architecture lays the foundation for all-optical engineering of isolated flat bands, optical quasicrystals, and valley-selective topological phases.

We obtain the final effective Hamiltonian [see Appendix~\ref{sec:S4}]:
\begin{equation}
    H_{\text{eff}}^\tau = \hbar v (\tau k_x \sigma_x + k_y \sigma_y) + \left[ \Delta + \tau V_0 M_F(\mathbf{r}) \right] \sigma_z, \label{eq:Hamlp}
\end{equation}
where $M_F(\mathbf{r})$ is the spatially varying moir\'e-like Floquet mass acting as a synthetic, staggered magnetic exchange field
\begin{equation}
    M_F(\mathbf{r}) = \sum_{i < j}^{3} (\hat{\mathbf{n}}_i \times \hat{\mathbf{n}}_j)_z \sin\left( (\mathbf{q}_i - \mathbf{q}_j) \cdot \mathbf{r} + (\phi_i - \phi_j) \right). \label{eq:MF_def}
\end{equation}
This expression demonstrates the critical role of the phase delays $\phi_j$. In the absence of phase offsets ($\phi_i = \phi_j = 0$), the geometric symmetry of the lattice forces the spatially varying optical mass to vanish identically at any position. Conversely, introducing a uniform phase gradient $\phi_j = (j-1)\frac{2\pi}{3}$ shifts the sine harmonics out of phase with the geometric intensity landscape, establishing a periodic superlattice of alternating $+M_F$ and $-M_F$ mass domains that locally break time-reversal symmetry (TRS) while globally preserving it.

\begin{figure}[t]
    \centering
    \includegraphics[width=\linewidth]{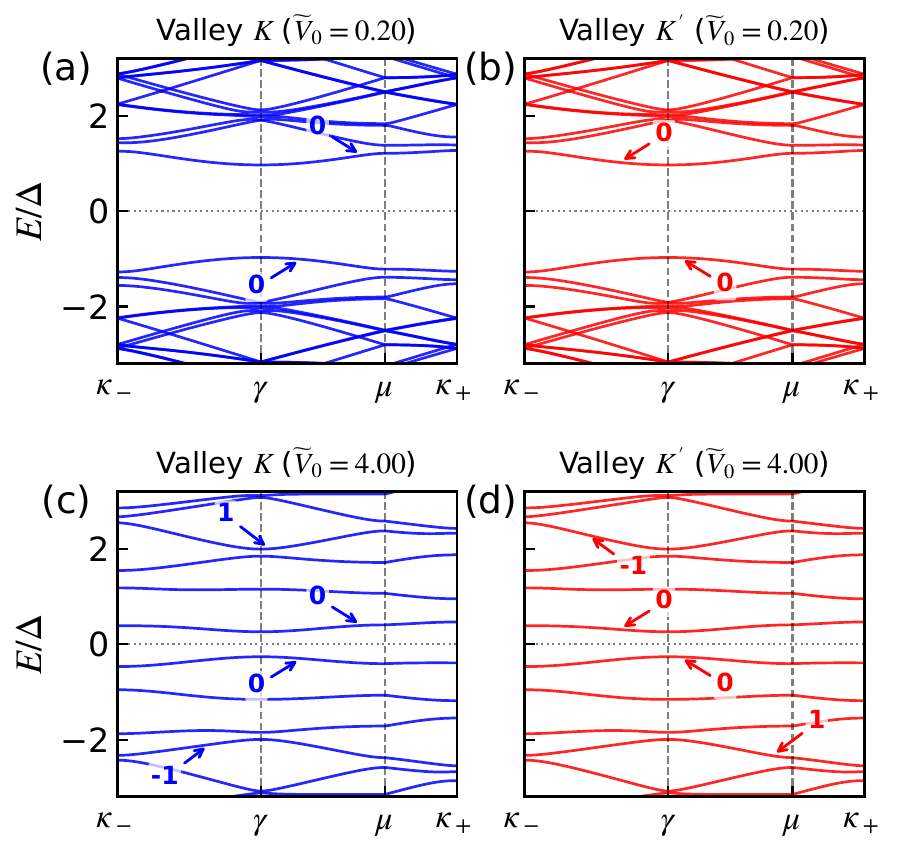}
    \caption{Valley-resolved Floquet moir\'e band structures of gapped Dirac fermions driven by $N=3$ phase-delayed linearly polarized (LP) beams for (a,c) Valley $K$ ($\tau=+1$) and (b,d) Valley $K'$ ($\tau=-1$) [see Fig.~\ref{fig:phasedelay}] with (a,b) $V_0=0.2\Delta $ and (c,d) $V_0=4 \Delta$.  The arrows indicate the  Chern numbers of the bands.}
    \label{fig:lpmoirebands}
\end{figure}

Figure~\ref{fig:lpmoirebands} shows the valley-resolved Floquet moir\'e minibands along the metric high-symmetry path $\boldsymbol{\kappa}_{-}\rightarrow\boldsymbol{\gamma}\rightarrow\boldsymbol{\mu}\rightarrow\boldsymbol{\kappa}_{+}$. Unlike the circularly polarized configuration discussed in Sec.~\ref{sec:2-cp}, the phase-delayed LP-SPP interference generates a spatially modulated Floquet mass with vanishing spatial average,
\begin{equation}
    \left\langle M_F(\mathbf{r}) \right\rangle = 0.
\end{equation}
Consequently, no valley-dependent mass correction is generated, and the $K$ and $K'$ valleys retain identical scalar spectra. The primary gap remains controlled predominantly by the intrinsic Dirac mass and is renormalized by the periodic scattering, yielding smaller gap for larger $V_0$ values [cf. Fig~\ref{fig:lpmoirebands}(a,b) vs (c,d)]. Away from $\boldsymbol{\gamma}$, the spatially modulated mass hybridizes the folded Dirac states and opens secondary moir\'e minigaps. Because the phase-shifted sine modulation breaks real-space inversion symmetry, $M_F(-\mathbf{r})\neq M_F(\mathbf{r})$, the individual miniband energies at $\boldsymbol{\kappa}_{-}$ and $\boldsymbol{\kappa}_{+}$ are not required to be identical.

The miniband topology is evaluated over the full mBZ using the Fukui-Hatsugai-Suzuki method~\cite{Fukui2005Chern}. 
For the parameters shown, the lowest conduction band and highest valence band are topologically trivial as indicated in Fig.~\ref{fig:lpmoirebands}. Under very strong field $V_0=4\Delta$, the adjacent isolated minibands nevertheless carry opposite valley-resolved Chern numbers, with the four valence bands following $C^{K}_v=(0,0,0,-1)$ and $C^{K'}_v=(0,0,0,+1)$ evaluated from the Fermi level [Fig.~\ref{fig:lpmoirebands}(c,d)]. Thus, although the band dispersions remain valley degenerate, their underlying quantum geometry is valley contrasting, satisfying
\begin{equation}
    C_n^{K'}=-C_n^{K}.
\end{equation}
The phase-delayed LP-SPP field therefore redistributes Berry curvature among the folded minibands without producing the macroscopic valley splitting characteristic of circularly polarized driving, yielding a compensated quantum-valley-Hall-like band topology.

\section{Conclusion}

In summary, we have proposed and theoretically demonstrated a purely optical Floquet framework to dynamically engineer moir\'{e}-like superlattices and quasicrystalline potentials in gapped Dirac materials. We established that off-resonant circularly polarized (CP) light generates a valley-dependent Floquet mass, providing a direct optical mechanism for controlling the Dirac spectrum. By designing an interference pattern of three coherent CP beams with a $120^\circ$ angular separation, we constructed a triangular mass superlattice that dynamically folds the Dirac spectrum into a mini-Brillouin zone, producing pronounced valley-selective miniband reconstruction and band flattening. Beyond this spectral reconstruction, we find that the valence miniband near the Fermi level undergoes a light-driven topological transition, with the transition confined to one valley within the parameter regime considered. Interestingly, by generalizing the interference to noncrystallographic multi-beam configurations, we extend the same mechanism beyond periodic superlattices and generate optical quasicrystalline potentials with long-range orientational order.

However, realizing these optical superlattices at deeply subwavelength length scales is constrained by the diffraction limit of free-space light. To overcome this limitation, we further consider highly confined surface plasmon polariton fields. Their multi-beam interference produces a spatially alternating chiral Floquet mass whose texture can be controlled through the relative optical phases. Because this chiral mass has zero spatial average, it does not generate the uniform valley-dependent mass shift characteristic of the CP configuration while still producing substantial miniband hybridization and secondary moir\'{e} minigaps. Crucially, although the $K$ and $K'$ valleys exhibit degenerate energy spectra, their corresponding minibands are topologically inequivalent and carry opposite valley-resolved Chern numbers, $C_n^{K'}=-C_n^K$. The plasmonic configuration therefore provides a distinct route to valley-contrasting topology without macroscopic valley splitting.

Ultimately, our findings establish a reconfigurable, all-optical platform for dynamically engineering moir\'{e} minibands, quasicrystalline electronic states, and valley-dependent topology without mechanical twisting. The ability to control the spatial symmetry, drive strength, and relative phases of the optical fields provides a versatile setting for exploring interaction effects and valley-selective phenomena in dynamically structured Dirac materials.

\begin{acknowledgments}
The authors acknowledge the MAHAMERU BRIN High Performance Computing (HPC) facility, provided by the National Research and Innovation Agency (BRIN), for the computational resources used in this work. We acknowledge the use of the Google Gemini large language model for language polishing and LaTeX formatting assistance during the preparation of this manuscript. The authors carefully reviewed and edited all AI-assisted content and take full responsibility for the originality, accuracy, and final integrity of this work.
\end{acknowledgments}
%

\bibliographystyle{apsrev4-2}
\clearpage
\onecolumngrid
\begin{center}
\textbf{\large Supplemental Material for: Floquet Engineering of moir\'e Flat Bands and Quasicrystals via Phase-Delayed Plasmonic Interference}
\end{center}
\vspace{0.5cm}

\setcounter{section}{0}
\setcounter{equation}{0}
\setcounter{figure}{0}

\renewcommand{\thesection}{S\arabic{section}}
\renewcommand{\theequation}{S\arabic{equation}}
\renewcommand{\thefigure}{S\arabic{figure}}

\section{Magnus Expansion and the Floquet Effective Hamiltonian}
\label{sec:gfactor}
The quantum dynamics of electrons in a Dirac material driven by a time-periodic electromagnetic field $H(t) = H(t + T)$ (with period $T = 2\pi/\omega$) is governed by the time-dependent Schr\"odinger equation:
\begin{equation}
    i\hbar \frac{\partial \psi(t)}{\partial t} = H(t) \psi(t) \implies \psi(t) = U(t, 0) \psi(0).
\end{equation}
The time-evolution operator $U(t, 0)$ satisfies the operator differential equation $\frac{d}{dt} U(t, 0) = -\frac{i}{\hbar} H(t) U(t, 0)$ subject to the initial condition $U(0, 0) = \mathbf{1}$. When the Hamiltonian at different times does not commute ($[H(t_1), H(t_2)] \neq 0$), repeatedly integrating this differential equation yields the formal Dyson series expansion:
\begin{equation}
    U(t, 0) = \mathbf{1} + \left(-\frac{i}{\hbar}\right) \int_0^t dt_1 H(t_1) + \left(-\frac{i}{\hbar}\right)^2 \int_0^t dt_1 \int_0^{t_1} dt_2 H(t_1) H(t_2) + \dots .
\end{equation}
Although the Dyson series provides an exact formal solution, truncating the series at any finite order generally breaks the unitarity ($U^\dagger U \neq \mathbf{1}$) of the evolution operator. To naturally preserve unitarity at all orders of approximation, Wilhelm Magnus \cite{Magnus1954Exponential} proposed representing $U(t,0)$ as a true single exponential:
\begin{equation}
    U(t, 0) = \exp\left( \Omega(t) \right) = \exp\left( \Omega_1(t) + \Omega_2(t) + \Omega_3(t) + \dots \right),
\end{equation}
where $\Omega(t)$ is an anti-Hermitian operator ($\Omega^\dagger = -\Omega$). Expanding $e^\Omega$ via Taylor series, $e^\Omega \approx \mathbf{1} + \Omega_1 + \Omega_2 + \frac{1}{2}\Omega_1^2$, and matching order-by-order against the Dyson series yields the first-order term ($\Omega_1$), which represents the time-averaged interaction Hamiltonian,
\begin{equation}
    \Omega_1(t) = -\frac{i}{\hbar} \int_0^t dt_1 H(t_1),
\end{equation}
and the second-order term ($\Omega_2$), which accumulates non-commutative time-fluctuations arising from $[H(t_1), H(t_2)]$:
\begin{equation}
    \Omega_2(t) + \frac{1}{2}\Omega_1^2(t) = \left(-\frac{i}{\hbar}\right)^2 \int_0^t dt_1 \int_0^{t_1} dt_2 H(t_1) H(t_2).
\end{equation}
Geometrically splitting the square integration domain $[0, t] \times [0, t]$ of $\frac{1}{2}\Omega_1^2(t)$ across the diagonal $t_1 = t_2$ into two symmetric triangles ($t_1 > t_2$ and $t_2 > t_1$) and exchanging dummy variables $t_1 \leftrightarrow t_2$ on the upper triangle gives:
\begin{equation}
    \frac{1}{2}\Omega_1^2(t) = \frac{1}{2} \left(-\frac{i}{\hbar}\right)^2 \int_0^t dt_1 \int_0^{t_1} dt_2 \left[ H(t_1)H(t_2) + H(t_2)H(t_1) \right].
\end{equation}
Substituting this back into the second-order equation directly isolates the commutator term:
\begin{equation}
    \Omega_2(t) = -\frac{1}{2\hbar^2} \int_0^t dt_1 \int_0^{t_1} dt_2 [H(t_1), H(t_2)].
\end{equation}
Because $H(t)$ is periodic in time ($H(t + T) = H(t)$), we decompose the Hamiltonian into a discrete Fourier series $H(t) = \sum_{m=-\infty}^{\infty} H_m e^{-im\omega t}$, where the Fourier components $H_m$ are evaluated via the integral projection:
\begin{equation}
    H_m = \frac{1}{T} \int_0^T H(t) e^{im\omega t} \, dt.
\end{equation}
The physical interpretation of each harmonic $H_m$ is straightforward: $H_0$ ($m = 0$) represents the static component; $H_{+1}$ ($m = +1$) carries the phase factor $e^{-i\omega t}$ and represents the quantum process of absorbing one photon quanta $\hbar\omega$; $H_{-1}$ ($m = -1$) carries the phase factor $e^{i\omega t}$ and represents emitting one photon quanta; and $H_m$ ($|m| > 1$) governs multi-photon transition processes.

According to Floquet's theorem, the stroboscopic evolution of the system over integer periods ($t = T, 2T, \dots$) can be mapped onto a static system governed by a time-independent Effective Hamiltonian ($H_{\text{eff}}$):
\begin{equation}
    U(T, 0) = \exp\left( -\frac{i}{\hbar} H_{\text{eff}} T \right) \implies H_{\text{eff}} = \frac{i\hbar}{T} \Omega(T) = H_{\text{eff}}^{(0)} + H_{\text{eff}}^{(1)} + \dots.
\end{equation}
Substituting the Fourier series into the Magnus operators and integrating over one full period $T$ yields the zeroth-order evaluation,
\begin{equation}
    \Omega_1(T) = -\frac{i}{\hbar} \sum_m H_m \int_0^T e^{-im\omega t_1} dt_1 = -\frac{i}{\hbar} H_0 T \implies H_{\text{eff}}^{(0)} = H_0,
\end{equation}
and the first-order evaluation,
\begin{equation}
    \Omega_2(T) = -\frac{1}{2\hbar^2} \sum_{m, n} [H_m, H_n] \int_0^T dt_1 \int_0^{t_1} dt_2 e^{-im\omega t_1} e^{-in\omega t_2}.
\end{equation}
The double time integral evaluates to non-zero values strictly under the resonance condition $n = -m$. Evaluating the integral and summing over positive harmonics $m \ge 1$ gives $\Omega_2(T) = \frac{i T}{\hbar^2 \omega} \sum_{m=1}^{\infty} \frac{[H_m, H_{-m}]}{m}$, yielding the first-order Floquet effective correction:
\begin{equation}
    H_{\text{eff}}^{(1)} = \frac{i\hbar}{T} \Omega_2(T) = -\frac{1}{\hbar\omega} \sum_{m=1}^{\infty} \frac{[H_m, H_{-m}]}{m} = \sum_{m=1}^{\infty} \frac{[H_{-m}, H_m]}{m\hbar\omega}.
\end{equation}

In the high-frequency off-resonant regime ($\hbar\omega \gg \{\Delta, E_F, V_{\text{int}}\}$), multi-photon contributions ($|m| > 1$) are strongly suppressed by factors of $1/m\hbar\omega$. Truncating the expansion at the fundamental harmonics ($m = \pm 1$) yields the canonical commutator formula:
\begin{equation}
    H_{\text{eff}}^\tau \approx H_0^\tau + \frac{[H_{-1}^\tau, H_1^\tau]}{\hbar\omega}.
\end{equation}
Physically, the commutator term $\frac{[H_{-1}, H_1]}{\hbar\omega}$ describes a second-order virtual photon-dressing process in which an electron absorbs a virtual photon ($H_1$) and subsequently re-emits it ($H_{-1}$). 

Now we consider a gapped Dirac material driven by a normally incident, uniform circularly polarized (CP) light with vector potential $\mathbf{A}(t) = A_0(\cos(\omega t)\hat{\mathbf{x}} + \eta \sin(\omega t)\hat{\mathbf{y}})$, where $\eta = \pm 1$ denotes optical helicity. Applying the Peierls substitution $\mathbf{p} \to \mathbf{p} - e\mathbf{A}(t)$ yields the interaction Hamiltonian $H_{\text{int}}^\tau(t) = -ev(\tau A_x(t)\sigma_x + A_y(t)\sigma_y)$. Decomposing $\cos(\omega t)$ and $\sin(\omega t)$ via Euler exponentials isolates the Fourier harmonics $m = \pm 1$, giving $H_1^\tau = -\frac{evA_0}{2}(\tau \sigma_x + i\eta \sigma_y)$ and $H_{-1}^\tau = -\frac{evA_0}{2}(\tau \sigma_x - i\eta \sigma_y)$. Evaluating the commutator using Pauli matrix algebra $[\sigma_x, \sigma_y] = 2i\sigma_z$ gives explicitly:
\begin{equation}
    [H_{-1}^\tau, H_1^\tau] = -e^2 v^2 A_0^2 \tau \eta \sigma_z.
\end{equation}
Substituting this commutator into $H_{\text{eff}}^\tau$ yields the dressed Floquet effective mass $\Delta_{\text{eff}}^\tau = \Delta - \frac{(evA_0)^2}{\hbar\omega}\tau\eta$, showing the optical Stark effect and dynamical valley splitting ($K$ vs $K'$).

The dynamical valley splitting induced by the photon dressing can be quantified by the energy difference between the two valleys:
\begin{equation}
    \Delta E_{\text{valley}} = 2\Delta_{\text{eff}}^{K} - 2\Delta_{\text{eff}}^{K'} = -4\frac{(evA_0)^2}{\hbar\omega} \eta.
\end{equation}
Physically, this valley splitting is equivalent to the application of a giant external magnetic field, a phenomenon analogous to the valley Zeeman effect. To formalize this analogy, we construct an effective pseudomagnetic field $B_{\text{eff}} \equiv \frac{e}{\hbar} A_0^2$ generated by the periodic driving. We can then factor the valley splitting equation into the standard Zeeman form, $\Delta E_{\text{valley}} = g_v \mu_B B_{\text{eff}}$, where $\mu_B = \frac{e\hbar}{2m_e}$ is the Bohr magneton. By equating the terms, we define the Floquet-engineered valley Land\'{e} $g$-factor:
\begin{equation}
    g_v = \frac{4 e v^2 / \omega}{e \hbar / 2m_e} = \frac{8 m_e v^2}{\hbar \omega}.
\end{equation}
This formulation reveals that the effective $g$-factor depends purely on the intrinsic properties of the material (via the Fermi velocity $v$) and the frequency of the external driving field $\omega$, completely independent of the laser intensity. For typical parameter values of $v=10^6\ \rm m/s$ and $\hbar \omega= 1\ \rm eV$, the valley $g$-factor is gigantic ($g=45.5$). The chirality of the light ($\eta$) serves as a directional switch, dictating which valley experiences an increase or decrease in its effective mass gap.
\section{Near Resonant (Sub Gap) Floquet Theory and Valley-Selective Optical Stark Effect}
\label{sec:resonant_floquet}

While the Floquet-Magnus expansion effectively captures the dynamical mass generation in the high-frequency limit ($\hbar\omega \gg 2\Delta_0$), describing the optical Stark effect in the near-resonant regime requires analyzing the extended Floquet Hilbert space. In this regime, the photon energy is slightly detuned below the unperturbed bandgap, defined by a small positive detuning parameter $\delta = 2\Delta_0 - \hbar\omega$.

We begin with the unperturbed Dirac Hamiltonian for a specific valley $\tau = \pm 1$, with an intrinsic static mass $\Delta_0$:
$$H_0^\tau = \hbar v (\tau k_x \sigma_x + k_y \sigma_y) + \Delta_0 \sigma_z$$
At the Dirac point ($\mathbf{k}=0$), the conduction band state is $|c\rangle = (1, 0)^T$ with energy $+\Delta_0$, and the valence band state is $|v\rangle = (0, 1)^T$ with energy $-\Delta_0$. 

The application of a circularly polarized driving field with chirality $\eta = \pm 1$ is introduced via the vector potential $\mathbf{A}(t) = A_0 (\cos(\omega t), \eta \sin(\omega t))$. The Peierls substitution yields the time-dependent interaction Hamiltonian:
$$H_{\text{int}}^\tau(t) = -e v A_0 (\tau \cos(\omega t) \sigma_x + \eta \sin(\omega t) \sigma_y)$$
By decomposing this into co-rotating and counter-rotating components ($H_{\text{int}}^\tau(t) = H_{+1}^\tau e^{-i\omega t} + H_{-1}^\tau e^{i\omega t}$), we isolate the photon absorption operator:
$$H_{+1}^\tau = -\frac{e v A_0}{2} \begin{pmatrix} 0 & \tau + \eta \\ \tau - \eta & 0 \end{pmatrix}$$

The near-resonant optical Stark effect is governed by the virtual absorption of a photon coupling the valence and conduction bands. The interband transition matrix element $M_{cv}^\tau$ evaluates to:
$$M_{cv}^\tau = \langle c | H_{+1}^\tau | v \rangle = -\frac{e v A_0}{2} (\tau + \eta)$$
The coupling strength is given by the squared magnitude of this matrix element. Since $\tau$ and $\eta$ are constrained to $\pm 1$, the expression simplifies algebraically:
$$|M_{cv}^\tau|^2 = (e v A_0)^2 \left( \frac{1 + \tau\eta}{2} \right) = (e v A_0)^2 \delta_{\tau, \eta}$$
This factor acts as a strict optical selection rule. The bands are strongly coupled ($|M_{cv}^\tau|^2 = (e v A_0)^2$) only when the optical chirality matches the valley index ($\tau = \eta$). Conversely, the transition is mathematically forbidden when $\tau = -\eta$.

To find the dressed energy levels, we project the system onto a $2 \times 2$ near-resonant Floquet subspace containing the zero-photon conduction band $|c, 0\rangle$ and the one-photon valence band $|v, 1\rangle$. The unperturbed energy of $|v, 1\rangle$ in this extended space is $-\Delta_0 + \hbar\omega = \Delta_0 - \delta$. The truncated Floquet Hamiltonian is:
$$\mathcal{H}_{\text{res}}^\tau = \begin{pmatrix} \Delta_0 & M_{cv}^\tau \\ (M_{cv}^\tau)^* & \Delta_0 - \delta \end{pmatrix}$$
Diagonalizing this matrix in the perturbative limit ($|M_{cv}^\tau| \ll \delta$) yields the dressed conduction band energy. The resulting optical Stark shift is:
$$\Delta E_{\text{shift}}^\tau = \frac{|M_{cv}^\tau|^2}{\delta} = \frac{(e v A_0)^2}{\delta} \delta_{\tau, \eta}$$
This confirms that the near-resonant optical Stark effect is perfectly valley-selective, explicitly breaking time-reversal symmetry by widening the gap in the coupled valley while leaving the opposite valley invariant \cite{Sie2015Valley}.

\subsection*{Comparison of Effective $g$-factors}
We can formalize the analogy to a magnetic field by defining a dynamical valley $g$-factor for both the near-resonant and off-resonant (Floquet-Magnus) limits. Using the Zeeman relation $\Delta E = g_v \mu_B B_{\text{eff}}$ with $B_{\text{eff}} = \frac{e}{\hbar}A_0^2$ and $\mu_B = \frac{e\hbar}{2m_e}$, we compare the two regimes.

In the high-frequency Floquet-Magnus limit ($\hbar\omega \gg 2\Delta_0$), the energy shift is inversely proportional to the driving frequency:
$$\Delta E_{\text{FM}}^\tau = \frac{(e v A_0)^2}{\hbar\omega} \quad \implies \quad g_{\text{FM}} = \frac{8 m_e v^2}{\hbar\omega}$$

In the near-resonant limit ($\hbar\omega \lesssim 2\Delta_0$), the energy shift is governed by the detuning $\delta$:
$$\Delta E_{\text{res}}^\tau = \frac{(e v A_0)^2}{\delta} \quad \implies \quad g_{\text{res}} = \frac{8 m_e v^2}{\delta}$$

The ratio of the effective $g$-factors perfectly illustrates the resonant enhancement:
$$\frac{g_{\text{res}}}{g_{\text{FM}}} = \frac{\hbar\omega}{\delta}$$
Because $\delta$ can be made arbitrarily small in a controlled experiment, the near-resonant $g$-factor ($g_{\text{res}}$) scales significantly larger than the off-resonant $g$-factor ($g_{\text{FM}}$). While the Floquet-Magnus approach provides the fundamental mechanism for generating a macroscopic topological mass, tuning the optical field near an excitonic resonance provides an experimental knob to amplify the synthetic magnetic field by orders of magnitude.

\section{Plane-Wave Expansion Matrix Elements}

To synthesize a two-dimensional spatial mass landscape dynamically, we generalize the periodic drive to $N$ coherent, circularly polarized (CP) plane waves intersecting symmetrically at the sample plane. The in-plane wavevector of the $j$-th beam is distributed symmetrically in the azimuthal plane:
\begin{equation}
    \mathbf{q}_j = q_0 \left( \cos\theta_j \hat{\mathbf{x}} + \sin\theta_j \hat{\mathbf{y}} \right) = q_0 \left[ \cos\left(\frac{2\pi j}{N}\right)\hat{\mathbf{x}} + \sin\left(\frac{2\pi j}{N}\right)\hat{\mathbf{y}} \right], \quad j = 1, 2, \dots, N.
\end{equation}
Where $q_0$ denotes the characteristic momentum scale of the incident driving field. The total vector potential in the sample plane is given by the linear superposition:
\begin{equation}
    \mathbf{A}(\mathbf{r}, t) = \sum_{j=1}^N \text{Re}\left[ A_0 (\hat{\mathbf{x}} + i\eta \hat{\mathbf{y}}) e^{i(\mathbf{q}_j \cdot \mathbf{r} - \omega t)} \right] = \frac{A_0}{2}(\hat{\mathbf{x}} + i\eta \hat{\mathbf{y}}) f_N(\mathbf{r}) e^{-i\omega t} + \text{c.c.},
\end{equation}
where $\eta = \pm 1$ dictates the optical helicity, and the complex spatial interference function is defined as:
\begin{equation}
    f_N(\mathbf{r}) \equiv \sum_{j=1}^N e^{i\mathbf{q}_j \cdot \mathbf{r}}.
\end{equation}

Coupling the electromagnetic field to low-energy Dirac electrons via the Peierls substitution $\mathbf{p} \to \mathbf{p} - e\mathbf{A}(\mathbf{r}, t)$ gives the interaction Hamiltonian $H_{\text{int}}^\tau(\mathbf{r}, t) = -ev(\tau A_x \sigma_x + A_y \sigma_y)$. In the high-frequency off-resonant regime ($\hbar\omega \gg \{\Delta, E_F, V_{\text{int}}\}$), the first-order Magnus expansion yields the effective Hamiltonian:
\begin{equation}
    H_{\text{eff}}^\tau(\mathbf{r}) \approx H_0^\tau + \frac{[H_{-1}^\tau(\mathbf{r}), H_1^\tau(\mathbf{r})]}{\hbar\omega} = v(\tau p_x \sigma_x + p_y \sigma_y) + \Delta_{\text{eff}}^\tau(\mathbf{r})\sigma_z.
\end{equation}
Evaluating the fundamental harmonic commutator $[H_{-1}^\tau(\mathbf{r}), H_1^\tau(\mathbf{r})] = -(evA_0)^2 \eta\tau |f_N(\mathbf{r})|^2 \sigma_z$ and expanding the spatial intensity profile $|f_N(\mathbf{r})|^2 = f_N(\mathbf{r})f_N^*(\mathbf{r})$ yields:
\begin{equation}
    |f_N(\mathbf{r})|^2 = \sum_{j=1}^N \sum_{k=1}^N e^{i(\mathbf{q}_j - \mathbf{q}_k)\cdot\mathbf{r}} = N + 2\sum_{j < k}^N \cos(\mathbf{b}_{jk}\cdot\mathbf{r}),
\end{equation}
where $\mathbf{b}_{jk} \equiv \mathbf{q}_j - \mathbf{q}_k$ defines the set of fundamental pairwise scattering vectors. Consequently, the spatially modulated Dirac mass landscape takes the general form:
\begin{equation}
    \Delta_{\text{eff}}^\tau(\mathbf{r}) = \Delta - \frac{(evA_0)^2}{\hbar\omega}\eta\tau \left[ N + 2\sum_{j < k}^N \cos(\mathbf{b}_{jk} \cdot \mathbf{r}) \right].
\end{equation}

The spatial periodicity of $\Delta_{\text{eff}}^\tau(\mathbf{r})$ depends critically on the beam number $N$. For periodic moir\'e superlattices ($N = 3, 4, 6$), the set of scattering vectors $\{\mathbf{b}_{jk}\}$ spans a two-dimensional Bravais lattice (triangular, square, and hexagonal, respectively). The system exhibits full discrete translational symmetry, rendering Bloch's theorem and the plane-wave expansion framework exact. Conversely, for optical quasicrystals ($N = 5, 7, \dots$), the configuration exhibits non-crystallographic rotational symmetries (5-fold, 7-fold) lacking long-range translational order. The reciprocal momentum space becomes infinitely dense, invalidating standard Brillouin zone truncation and necessitating real-space polynomial expansions (e.g., the Kernel Polynomial Method).

For periodic configurations, we partition $\Delta_{\text{eff}}^\tau(\mathbf{r})$ into a uniform background mass shift $m_0^\tau$ and a scattering potential modulation amplitude $V_M^\tau$:
\begin{equation}
    m_0^\tau \equiv \Delta - N \frac{(evA_0)^2}{\hbar\omega}\eta\tau, \quad V_M^\tau \equiv -\frac{(evA_0)^2}{\hbar\omega}\eta\tau.
\end{equation}
Expanding $2\cos(\mathbf{b}_{jk}\cdot\mathbf{r}) = e^{i\mathbf{b}_{jk}\cdot\mathbf{r}} + e^{-i\mathbf{b}_{jk}\cdot\mathbf{r}}$ across the shell of primary moir\'e reciprocal lattice vectors $G_1 = \{\pm \mathbf{b}_{jk}\}$, the Floquet mass landscape casts into a compact Fourier series:
\begin{equation}
    \Delta_{\text{eff}}^\tau(\mathbf{r}) = m_0^\tau + V_M^\tau \sum_{\mathbf{g} \in G_1} e^{i\mathbf{g} \cdot \mathbf{r}}.
\end{equation}

Owing to the spatial periodicity $\Delta_{\text{eff}}^\tau(\mathbf{r} + \mathbf{R}_M) = \Delta_{\text{eff}}^\tau(\mathbf{r})$, the electronic state with crystal momentum $\mathbf{k}$ inside the miniature moir\'e Brillouin Zone (mBZ) is expanded onto a plane-wave spinor basis:
\begin{equation}
    \Psi_{\mathbf{k}}^\tau(\mathbf{r}) = \frac{1}{\sqrt{\mathcal{A}}} \sum_{\mathbf{G}} e^{i(\mathbf{k} + \mathbf{G})\cdot\mathbf{r}} \mathbf{c}_{\mathbf{G}} = \frac{1}{\sqrt{\mathcal{A}}} \sum_{\mathbf{G}} e^{i(\mathbf{k} + \mathbf{G})\cdot\mathbf{r}} \begin{pmatrix} c_{A, \mathbf{G}} \\ c_{B, \mathbf{G}} \end{pmatrix},
\end{equation}
where $\mathcal{A}$ is the normalization area, $\mathbf{G} = n_1 \mathbf{b}_1 + n_2 \mathbf{b}_2$ ($n_1, n_2 \in \mathbb{Z}$) denotes the 2D moir\'e reciprocal lattice vectors, and $\mathbf{c}_{\mathbf{G}}$ is the two-component spinor amplitude in sublattice space $(A, B)$.

Substituting the ansatz into the stationary eigenvalue equation $H_{\text{eff}}^\tau \Psi_{\mathbf{k}}^\tau(\mathbf{r}) = E_{\mathbf{k}} \Psi_{\mathbf{k}}^\tau(\mathbf{r})$ and evaluating the differential momentum operator $\mathbf{p} = -i\hbar\boldsymbol{\nabla}$ gives:
\bea
    \frac{1}{\sqrt{\mathcal{A}}} \sum_{\mathbf{G}} \left[ \hbar v \big( \tau (k_x + G_x)\sigma_x + (k_y + G_y)\sigma_y \big) + m_0^\tau \sigma_z \right] e^{i(\mathbf{k}+\mathbf{G})\cdot\mathbf{r}} \mathbf{c}_{\mathbf{G}} + \frac{1}{\sqrt{\mathcal{A}}} \sum_{\mathbf{G}} \sum_{\mathbf{g} \in G_1} V_M^\tau \sigma_z e^{i(\mathbf{k}+\mathbf{G}+\mathbf{g})\cdot\mathbf{r}} \mathbf{c}_{\mathbf{G}} &&=\nn
    E_{\mathbf{k}} \frac{1}{\sqrt{\mathcal{A}}} \sum_{\mathbf{G}} e^{i(\mathbf{k}+\mathbf{G})\cdot\mathbf{r}} \mathbf{c}_{\mathbf{G}}&&.
\eea
Projecting this equation onto the dual basis state by multiplying from the left with $\frac{1}{\sqrt{\mathcal{A}}} e^{-i(\mathbf{k}+\mathbf{G}')\cdot\mathbf{r}}$ and integrating over real space $\int_{\mathcal{A}} d^2\mathbf{r}$, the orthogonality relation $\frac{1}{\mathcal{A}}\int_{\mathcal{A}} d^2\mathbf{r} \, e^{i(\mathbf{Q}-\mathbf{Q}')\cdot\mathbf{r}} = \delta_{\mathbf{Q}', \mathbf{Q}}$ transforms the continuous differential equation into an infinite-dimensional coupled algebraic matrix eigenvalue problem:
\begin{equation}
    \sum_{\mathbf{G}} \mathcal{H}_{\mathbf{G}', \mathbf{G}}^\tau(\mathbf{k}) \mathbf{c}_{\mathbf{G}} = E_{\mathbf{k}} \mathbf{c}_{\mathbf{G}'}.
\end{equation}

The $2 \times 2$ block matrix elements $\mathcal{H}_{\mathbf{G}', \mathbf{G}}^\tau(\mathbf{k})$ couple different momentum channels and decompose into kinetic and scattering contributions. The diagonal kinetic elements ($\mathbf{G}' = \mathbf{G}$) represent the unperturbed massless Dirac dispersion evaluated at the shifted momentum $\mathbf{k} + \mathbf{G}$, shifted by the uniform background mass:
\begin{equation}
    \mathcal{H}_{\mathbf{G}, \mathbf{G}}^\tau(\mathbf{k}) = \hbar v \big[ \tau (k_x + G_x)\sigma_x + (k_y + Gy)\sigma_y \big] + m_0^\tau \sigma_z
\end{equation}
\begin{equation}
    \mathcal{H}_{\mathbf{G}, \mathbf{G}}^\tau(\mathbf{k}) = \hbar v \begin{pmatrix} \frac{m_0^\tau}{\hbar v} & \tau (k_x + G_x) - i(k_y + G_y) \\ \tau (k_x + G_x) + i(k_y + G_y) & -\frac{m_0^\tau}{\hbar v} \end{pmatrix}.
\end{equation}

The off-diagonal scattering elements ($\mathbf{G}' \neq \mathbf{G}$) arise because the periodic Floquet potential acts as a quantum diffraction grating. Inter-state scattering occurs exclusively when the momentum transfer matches a primary reciprocal vector ($\mathbf{G}' - \mathbf{G} = \mathbf{g} \in G_1$). For circularly polarized light, these off-diagonal blocks are purely real:
\begin{equation}
    \mathcal{H}_{\mathbf{G}+\mathbf{g}, \mathbf{G}}^\tau = V_M^\tau \sigma_z = \begin{pmatrix} V_M^\tau & 0 \\ 0 & -V_M^\tau \end{pmatrix}.
\end{equation}
For any other momentum mismatch ($\mathbf{G}' - \mathbf{G} \notin G_1$), the coupling vanishes identically: $\mathcal{H}_{\mathbf{G}', \mathbf{G}}^\tau = \mathbf{0}$.

To compute the moir\'e miniband dispersion numerically, the infinite plane-wave basis is truncated to a finite subspace satisfying $|\mathbf{G}| \le G_{\text{max}}$. This truncation is physically justified since the Dirac kinetic energy scales linearly with momentum ($E_{\text{kin}} \sim \hbar v |\mathbf{k} + \mathbf{G}|$), whereas the scattering matrix element $V_M^\tau$ remains constant. Consequently, the perturbative coupling ratio to high-momentum states decays asymptotically as $|V_M^\tau|/(\hbar v |\mathbf{G}|) \to 0$. Truncating at the 3rd or 4th shell of the mBZ yields a well-converged basis dimension $N_G \sim 30\text{-}100$. Repeated diagonalization of the resulting $2N_G \times 2N_G$ Hermitian matrix $\mathcal{H}^\tau(\mathbf{k})$ along the high-symmetry path ($\kappa \to \gamma \to \mu \to \kappa$) yields the exact Floquet-Moir\'e band structure and captures the emergence of isolated flat bands at the magic optical drive amplitude.




\section{Effective Floquet Hamiltonian for $N=3$ Phase-Delayed Beams \label{sec:S4}}

We begin with the low-energy effective Dirac Hamiltonian for a two-dimensional material, evaluated at the valleys $\tau = \pm 1$, which includes a static mass gap $\Delta$:
\begin{equation}
    H_0 = \hbar v (\tau k_x \sigma_x + k_y \sigma_y) + \Delta \sigma_z
\end{equation}
The light-matter interaction is introduced via the minimal coupling substitution $\mathbf{k} \rightarrow \mathbf{k} + \frac{e}{\hbar} \mathbf{A}(\mathbf{r}, t)$. The time-dependent Hamiltonian is thus given by:
\begin{equation}
    H(\mathbf{r}, t) = H_0 + e v \tau (\boldsymbol{\sigma} \cdot \mathbf{A}(\mathbf{r}, t))
\end{equation}

For the interference of $N=3$ intersecting, linearly polarized beams with a specific phase delay $\phi_j$, the total vector potential is the superposition of the individual waves:
\begin{equation}
    \mathbf{A}(\mathbf{r}, t) = \sum_{j=1}^{3} A_0 \hat{\mathbf{u}}_j \cos(\mathbf{q}_j \cdot \mathbf{r} - \omega t + \phi_j)
\end{equation}
where $\hat{\mathbf{u}}_j$ is the linear polarization vector (transverse or longitudinal), $\mathbf{q}_j$ is the in-plane wavevector, and $\phi_j$ represents the controlled phase delay (e.g., $\phi_1=0$, $\phi_2=\frac{2\pi}{3}$, $\phi_3=\frac{4\pi}{3}$).

We express this vector potential in terms of its time-harmonic Fourier components:
\begin{equation}
    \mathbf{A}(\mathbf{r}, t) = \mathbf{A}_{+1}(\mathbf{r}) e^{-i\omega t} + \mathbf{A}_{-1}(\mathbf{r}) e^{i\omega t}
\end{equation}
where the spatial components are $\mathbf{A}_{+1}(\mathbf{r}) = \frac{A_0}{2} \sum_{j=1}^3 \hat{\mathbf{u}}_j e^{i(\mathbf{q}_j \cdot \mathbf{r} + \phi_j)}$ and $\mathbf{A}_{-1} = \mathbf{A}_{+1}^*$. This yields the time-dependent perturbation harmonics:
\begin{equation}
    H_{\pm 1}(\mathbf{r}) = \tau \frac{e v A_0}{2} \sum_{j=1}^{3} (\boldsymbol{\sigma} \cdot \hat{\mathbf{u}}_j) e^{\pm i(\mathbf{q}_j \cdot \mathbf{r} + \phi_j)}
\end{equation}

In the high-frequency limit, where the photon energy $\hbar\omega$ is significantly larger than the relevant electronic energy scales, the system dynamics are well-described by the time-independent effective Hamiltonian up to the first order of the Magnus expansion:
\begin{equation}
    H_{\text{eff}} = H_0 + \frac{[H_{-1}, H_{+1}]}{\hbar \omega}
\end{equation}

To evaluate the commutator, we apply the Pauli matrix identity $[\boldsymbol{\sigma} \cdot \mathbf{a}, \boldsymbol{\sigma} \cdot \mathbf{b}] = 2i \boldsymbol{\sigma} \cdot (\mathbf{a} \times \mathbf{b})$. Given that the polarization vectors $\hat{\mathbf{u}}_j$ lie entirely within the two-dimensional $xy$-plane, their cross product is strictly aligned with the $z$-direction:
\begin{equation}
    [\boldsymbol{\sigma} \cdot \hat{\mathbf{u}}_i, \boldsymbol{\sigma} \cdot \hat{\mathbf{u}}_j] = 2i (\hat{\mathbf{u}}_i \times \hat{\mathbf{u}}_j)_z \sigma_z
\end{equation}

Expanding the commutator and applying Euler's formula to the complex exponentials yields the final effective Hamiltonian:
\begin{equation}
    H_{\text{eff}} = \hbar v (\tau k_x \sigma_x + k_y \sigma_y) + \left[ \Delta + \tau M_F(\mathbf{r}) \right] \sigma_z
\end{equation}
where $M_F(\mathbf{r})$ is the spatially varying moir\'e Floquet mass, which acts as a synthetic, staggered magnetic exchange field:
\begin{equation}
    M_F(\mathbf{r}) = \frac{(e v A_0)^2}{\hbar \omega} \sum_{i < j}^{3} (\hat{\mathbf{u}}_i \times \hat{\mathbf{u}}_j)_z \sin\left( (\mathbf{q}_i - \mathbf{q}_j) \cdot \mathbf{r} + (\phi_i - \phi_j) \right)
\end{equation}

This expression demonstrates the critical role of the phase delays $\phi_j$. In the absence of phase delays ($\phi_i = \phi_j = 0$), the geometric symmetry of the lattice forces the spatially varying optical mass to average to zero, producing no net local circular polarization. Conversely, non-zero phase offsets (such as $\phi_j = (j-1)\frac{2\pi}{3}$) shift the sine waves out of phase with the geometric intensity lattice, establishing a periodic moir\'e superlattice of alternating $+M_F$ and $-M_F$ mass domains that locally break time-reversal symmetry but globally preserve it.

\section{Numerical Evaluation of Chern Numbers  \label{sec:S5}}

To evaluate the topological invariants on a discretized moir\'e Brillouin zone (mBZ) without numerical gauge ambiguities, we employ the lattice gauge formulation developed by Fukui, Hatsugai, and Suzuki (FHS). The mBZ is discretized onto a uniform $N_k \times N_k$ mesh on the closed torus $T^2$:
\begin{equation}
    \mathbf{k} = s_1 \mathbf{b}_1 + s_2 \mathbf{b}_2, \quad s_{1,2} \in \left[-\frac{1}{2}, \frac{1}{2}\right)
\end{equation}
with grid steps $\Delta \mathbf{k}_1 = \frac{\mathbf{b}_1}{N_k}$ and $\Delta \mathbf{k}_2 = \frac{\mathbf{b}_2}{N_k}$.

To dynamically switch between isolated single-band (Abelian) and degenerate multiplet (Non-Abelian) formulations, we compute the global minimum direct gap between adjacent bands across the 2D torus:
\begin{equation}
    \delta_n = \min_{\mathbf{k} \in T^2} \left[ E_{n+1}(\mathbf{k}) - E_n(\mathbf{k}) \right]
\end{equation}
\begin{itemize}
    \item If $\delta_n \ge \delta_{\text{threshold}}$, the $n$-th band is isolated and evaluated using the Abelian FHS scheme ($M=1$).
    \item If $\delta_n < \delta_{\text{threshold}}$, local band touching occurs, and adjacent states are clustered into a composite subspace $\mathcal{S} = \{n, n+1, \dots, n+M-1\}$ of dimension $M \ge 2$, requiring the Non-Abelian FHS formulation.
\end{itemize}
When crossing the mBZ boundary, the plane-wave spinor basis functions transform according to:
\begin{equation}
    \Psi(\mathbf{k} + \mathbf{b}_\mu) = \mathcal{S}_{\mathbf{b}_\mu} \Psi(\mathbf{k}), \quad (\mu \in \{1, 2\})
\end{equation}
where $\mathcal{S}_{\mathbf{b}_\mu}$ is the unitary basis-shift operator defined by $(\mathcal{S}_{\mathbf{b}_\mu})_{\mathbf{G},\mathbf{G}'} = \delta_{\mathbf{G} + \mathbf{b}_\mu, \mathbf{G}'} \otimes \mathbb{I}_{2\times 2}$.

For an $M$-dimensional manifold spanned by $\Psi(\mathbf{k}) = \left[ |u_1(\mathbf{k})\rangle, \dots, |u_M(\mathbf{k})\rangle \right] \in \mathbb{C}^{2N_G \times M}$, the $M \times M$ non-Abelian overlap matrices along the primitive directions are:
\begin{equation}
    \mathcal{M}_1(\mathbf{k}) = \Psi^\dagger(\mathbf{k}) \Psi(\mathbf{k} + \Delta\mathbf{k}_1), \quad \mathcal{M}_2(\mathbf{k}) = \Psi^\dagger(\mathbf{k}) \Psi(\mathbf{k} + \Delta\mathbf{k}_2)
\end{equation}
The gauge-invariant lattice field strength $\tilde{F}_{12}(\mathbf{k})$ on each elementary plaquette is evaluated via the closed Wilson loop operator $\mathcal{W}_{\text{plaq}}(\mathbf{k})$:
\begin{equation}
    \mathcal{W}_{\text{plaq}}(\mathbf{k}) = \mathcal{M}_1(\mathbf{k}) \mathcal{M}_2(\mathbf{k} + \Delta\mathbf{k}_1) \mathcal{M}_1^\dagger(\mathbf{k} + \Delta\mathbf{k}_2) \mathcal{M}_2^\dagger(\mathbf{k})
\end{equation}
\begin{equation}
    \tilde{F}_{12}(\mathbf{k}) = \text{Arg}\left( \det \mathcal{W}_{\text{plaq}}(\mathbf{k}) \right) \in (-\pi, \pi]
\end{equation}
For an isolated single band ($M=1$), this expression reduces to the standard scalar $U(1)$ link variable product:
\begin{equation}
    \mathcal{W}_{\text{plaq}}^{(M=1)}(\mathbf{k}) = U_1(\mathbf{k}) U_2(\mathbf{k} + \Delta\mathbf{k}_1) U_1^*(\mathbf{k} + \Delta\mathbf{k}_2) U_2^*(\mathbf{k})
\end{equation}
where
\begin{equation}
    U_\mu(\mathbf{k}) = \frac{\langle u(\mathbf{k}) | u(\mathbf{k} + \Delta\mathbf{k}_\mu) \rangle}{|\langle u(\mathbf{k}) | u(\mathbf{k} + \Delta\mathbf{k}_\mu) \rangle|}
\end{equation}
The total Chern number $C_{\mathcal{S}}$ of the manifold is the sum of the lattice field strength over all $N_k^2$ plaquettes on the closed torus:
\begin{equation}
    C_{\mathcal{S}} = \frac{1}{2\pi} \sum_{\mathbf{k} \in T^2} \tilde{F}_{12}(\mathbf{k}) \in \mathbb{Z}
\end{equation}
This formulation ensures that $C_{\mathcal{S}}$ is strictly quantized to an integer, fully immune to gauge choices or internal band touching within the target manifold.

\end{document}